# The impact of varying electrical stimulation parameters on neuromuscular response

Dhruv Pai, Mentor Kip Ludwig

**Abstract:**

High density neurostimulation systems are coming to market to help spinal cord injury patients by stimulating and recording neuromuscular function. However, the parameter space that these systems have to explore is exceedingly large, and would need an artificial intelligence (AI) system to optimize. We need a platform that will allow us to determine the optimal parameter space for these systems. Our project aims to build a platform for mapping and controlling neuromuscular activity, as a high-throughput testbed for implementing and testing closed-loop neuromuscular activity. This abstract presents the first phase (the mapping phase) of building that testbed by combining multi-electrode stimulation/recording with visual motion-tracking. A 3D-printed rectangular raceway was used with 4 pairs of differential recording electrodes, and two stimulation electrodes embedded in the raceway bed. Non-anesthetized earthworms were placed on the raceway with their head section on the stimulating electrodes. Bipolar sinusoidal stimulation pulses of a range of voltages (2 to 6Vp-p), pulse durations (2 ms to 6.7 ms), and a burst rate of 1 pulse per second were applied, and action potentials and physical motion were recorded and analyzed. Action potentials were found to correlate with expansion/contraction displacements of worm segments, and voltage increases were shown to increase action potential propagation amplitude. Using the multiple electrode recording allowed us to capture the wave propagation of action potential pulse over the length of the worm. Feasibility of a platform to simultaneously monitor action potentials and motion of earthworms with real-time mapping was demonstrated.

**Introduction:**

According to the World Health Organization 2013 report, people with spinal cord injury (SCI) are two-to-five times more likely to die prematurely than people without a SCI, with worse survival rates in low- and middle-income countries. SCI symptoms may include partial or complete loss of sensory function or motor control of the arms, legs, and/or body.

High density neurostimulation systems are coming to market to help SCI patients by stimulating and recording neuromuscular function. However, the parameter space that these systems have to explore is exceedingly large, and would need an artificial intelligence (AI) system to optimize. We need a testbed that will allow us to determine the optimal parameter space for these systems. My project aims to build NeuroMACS, a system for mapping and controlling neuromuscular activity, as a high-throughput testbed for implementing and testing closed-loop neuromuscular activity.

This report presents the first phase (the mapping phase) of building that testbed, and combines multi-electrode stimulation/recording with motion tracking.

**Background:**

According to the World Health Organization 2013 statistics, every year, around the world, between 250,000 and 500,000 people suffer a spinal cord injury (SCI). The National SCI Statistical Center states that as of 2016, the annual incidence of SCI in the US is approximately 54 cases per million population. While no universal cure for SCI exists, various forms of stimulation systems have been marketed to tackle the needs related to SCI. Epidural spinal cord stimulation (SCS), using a surgically implanted stimulator in the epidural space of the spine, yields focused stimulation, but has downsides of increased hospital visits, increased medical expenses, and potential internal infections. Transcutaneous electrical nerve stimulators (TENS), on the other hand, focus on using transcutaneous stimulation to provide pain relief close to the surface of the skin. While TENS is used to stimulate sensory nerves, Functional Electrical Stimulation (FES), an approach similar to TENS, stimulates motor nerves to enable muscle contraction.While FES, like TENS, is easier than SCS and needs no clinical assistance to set up, controlling which nerve branch gets affected is difficult to do; this has impacted the ability to have reliable and reproducible treatments (Clark et al., 2007; Ragnarsson, 2007; Phillips, 2017).

Ever since Famm, et al. (2013) raised the idea of harnessing the electrical impulses of the nervous system to treat various diseases and to maintain healthy conditions, electroceuticals has been a field of active interest. As Famm, et al. indicate, virtually all organs and functions are regulated by neuronal circuits which form discrete components and which are controlled by specific patterns of electrical impulses ("action potentials"). This would suggest the possibility of developing a new approach for treating various diseases by modulating the electrical signaling patterns of the peripheral nervous system. Recently, Birmingham, et al. (2014) put forward a roadmap for realizing this vision. One of the key components of this roadmap is to decode neural signalling patterns that control individual organs, and the authors suggest that this will require "simultaneous recordings of neural signal and biomarkers of organ function (for example, blood pressure and cytokine release) that can be mined for correlations, and on stimulation and blocking experiments to test causation."

I envision building a system to map the effect of electrical stimulation on neuromuscular function and to control neuromuscular function using this mapping, with the ultimate goal to develop tools/techniques to help people who have spinal cord injury. With the advent of high-density electrode arrays for recording and stimulation of neuromuscular activity, I believe that such a system can become a closed-loop neuromuscular activation testbed utilizing an AI approach to loop over the entire parameter space of electrode pairs, pulse amplitude, pulse width, temporal patterning, etc, to determine the optimal parameter space for desired neuromuscular response.

This report outlines the current status of the first component in the development of this mapping and controlling system. NeuroMAPS: Neuromuscular Activity Mapping System is a system to map the effect of electrical stimulation on neuromuscular function, as measured by physical movement. The goal is to develop a map of the chronaxie times for different settings of the stimulation parameters, and correlate that to physical movements. Chronaxie time is the time required for an electric current with double the strength of the rheobase to stimulate a nerve fiber (Irnich, 1980). Rheobase is the minimum current that will produce an action potential or muscle contraction (Irnich, 1980).

To prototype and test NeuroMAPS, I plan to use earthworms as model organisms. Annelids, such as the common earthworms (*lumbricus terrestris*), and also arthropods, have a ventral nerve cord that is analogous to the dorsal spinal cord of vertebrates. The simple structure of the earthworm's nervous system makes it an ideal and inexpensive model organism to study ways to control physical movement by manipulating the nervous system. The skin and body muscle walls of the earthworm are thin; this means that electrical stimulation of, and recordings from, nerve fibers can be easily performed without dissection or surgical interventions. This means unsedated worms can be used for the study, and the worms will survive the experiment intact. Some of the early work in tracking the action potentials in earthworms had shown that while the medial giant fibre (MGF) conducts in the anterior posterior direction, the lateral giant fibre (LG) conducts in the reverse direction (with the rates being 32.2:12.6 m/s respectively) (Drewes, Landa, & McFall, 1978). Drewes, et al. also showed that longitudinal contractions of the earthworms occurred only when the MGF spikes were two or more, or when the LGF spikes were greater than two. Their work did not focus on the quantitative control of the stimulus strength, and did not map the muscular movement with changes in action potentials.

**Materials and Methods**

Worms were acquired from a local bait shop. The proof-of-concept version of the racetrack was a balsa wood-based rectangular racetrack, with 4 pairs of differential recording electrodes (individual electrode spacing of 1 cm). The stimulation section of the rectangular racetrack was made of plastic rather than balsa wood, and two stimulation electrodes were driven by a HP function generator.

Unsedated worms were placed on the racetrack with their head section on stimulating electrodes. Bipolar sinusoidal pulses of different pulse durations and amplitudes were used in a burst mode (1 pulse per second) for stimulation, and action potentials and physical motion were tracked. Signal from recording channels was fed into a 8-channel ASC902HF pre-amplifier (Astro-Med, Inc) and then fed into the ADI DAQ (Power Lab 8/30, AD Instruments Pty Ltd). An overhead iPhone XR was used to record real-time motion. During trials, qualitative motion observations and waveforms on an oscilloscope were used to adjust parameters of the function generator to enable better stimulation. All recording channel data was acquired on Chart software and post-processed in R. A mains filter and a low pass filter of 50 Hz were applied digitally to filter out AC and stimulation artifacts. Motion tracking was done using Tracker software (https://physlets.org/tracker; ©Douglas Brown). Vector lengths from worm head-to-midsection (HtM) and tail-to-midsection (TtM) were measured to track worm deformation over the duration of the stimulation.

**Results and Discussion**

Three distinct parameter spaces were considered in this phase: varying the stimulation pulse amplitude while keeping pulse width constant, varying the pulse width while keeping the pulse amplitude constant, and combining high pulse amplitude with long pulse width.

- <u>Varying pulse amplitudes while keeping the stimulation pulse width constant(n=2):</u>

Action potentials were observed (figure 2 and 3) when stimulation pulses of 3.33 ms pulse width were applied, with varying pulse amplitudes (2, 4, 5, and 6 $V_{p-p}$). Figures 2 and 3 (5V and 6Vp-p, respectively) show that as the stimulation amplitude increases, higher muscular contractions are observed as shown by extrema in displacement of body segments. Higher voltage also created greater amplitude of action potential propagation with all receiver channels observing higher amplitude of initial propagation in 6V sample relative to 5V sample. Note that channel 4 shows the reverse in polarity of action potential due to the conduction in the lateral giant fiber.

- <u>Varying the stimulation pulse width while keeping the pulse amplitude constant (n=2):</u>

The pulse amplitude was fixed at 4Vp-p, and the bipolar sinusoidal pulse width was varied: 2, 2.5, 2.86, 3.33, 4, 5, and 6.67 ms. Action potential recordings were obtained at both 5 ms and 6.67ms. For 6.67 ms pulses, both HtM and TtM vector lengths were tracked. For other samples, only the HtM could be tracked as the tail-end moved outside the field-of-view. Analysis of these two vectors in the 6.67 ms pulse recording showed that the worm experienced one large contraction immediately after stimulation. However, at 5 ms pulse width stimulation, the worm experienced a prolonged expansion along with one minor contraction. The worm also did not cover channel 4 for the 5ms recording leading to a lack of signal on that channel. Action potentials for 6.67ms pulse widths appeared to be better defined with channels 1, 2, and 3 approximating the same waveform. A propagation artifact can be seen occurring approximately

120 ms after stimulation. This artifact can model the wave propagation as it is observed in greatest amplitude at 120ms after stimulation in channel 1. It then appears at 150 and 180 ms in channels 2 and 3 respectively with lesser amplitude. Channel 4 showed no artifact, as it was the farthest from stimulation. A similar artifact occurred for the stimulation pulse width of 5ms. It appears in Channel 1 about 100 ms after stimulation and with lesser amplitude 130 ms after stimulation in channel 2. The setup has thus demonstrated capacity to measure wave propagation through 4 channels.

- <u>High pulse amplitude, and long pulse width (n=1)</u>:

Results were also acquired for stimulation pulses with 10 ms pulse width and pulse amplitude of 11 Vp-p, and appear to yield consistent signal recordings. Two separate action potentials appear to be present at 68 and 98 seconds in the recordings. Stimulation caused simultaneous oscillatory movement of the HtM and TtM sections for the worm. Interestingly, contraction of head-to-midsection segment was correlated to expansion of tail-to-midsection segment and vice versa.

    However, several design flaws and artifacts were noted during the development of this prototype. The first version of the prototype had both stimulating and recording electrodes embedded in the balsa wood. However, the wood would get damp, causing stimulation current to leak into the recording electrode signal, leading to significant stimulation artifacts. In order to eliminate this artifact, a plastic section was introduced between the balsa wood and the stimulation electrodes, and the recording shown in this report were obtained using this approach. However, there are still concerns about the electrical isolation of the various recording electrodes. In order to eliminate the testbed as a source of error, the second generation of the

prototype is a 3D-printed raceway (shown in Figure 7). Silver electrodes were inserted on the floor of the raceway so that the worms could ride on the electrodes (figures 7 and 8). There was also some concern about whether the signals observed in Figures 2-6 were primarily the stimulation pulse being tracked by the recording electrodes (i.e. the stimulation artifact) versus actual action potentials (i.e. was the electronics really amplifying the action potential signals to detectable amplitudes?). To address this concern, non-electrically-stimulated worms were evaluated on the new 3D-printed raceway, and evoked action potential signal was recorded for the 4 receiving channels. To minimize artifacts due to motion, a sealed beaker containing a worm immersed in water was cooled down by placing in a freezer for 20 minutes. The worm was then removed from the beaker and placed on the raceway (figure 8) and evoked action potentials were tracked for gentle tapping of the worm's body. Figure 9 shows that the action potentials are captured for three of the four recording electrode pairs (electrode pair 4 has very low signal). The signal is strongest at the first electrode pair, close to the head of the worm, and fades as we move towards the tail of the worm. This demonstrates that our electronics is able to detect the evoked action potentials in the earthworm (although the signal-to-noise could definitely use some improvement).

**Conclusions**:

The report successfully demonstrates the feasibility of NeuroMAPS, a system to simultaneously monitor action potentials and motion of unsedated earthworms with real-time mapping was demonstrated. The action potential recordings appear to suggest correlation between observed electrical signals and expansion/contraction of worm segments in motion recording. Increase in the amplitude of the stimulation pulses appears to increases the amplitude of the observed electrical signal, though the role of stimulation artifacts needs to be separated out. Using the multiple electrode recording enables the capture of wave propagation of action potential pulse over the length of the worm. For the case of long stimulation pulse width, the head-to-mid contractions of the worm body appears to correlate with mid-to tail expansions and vice versa, though more worm studies are required to validate this finding.

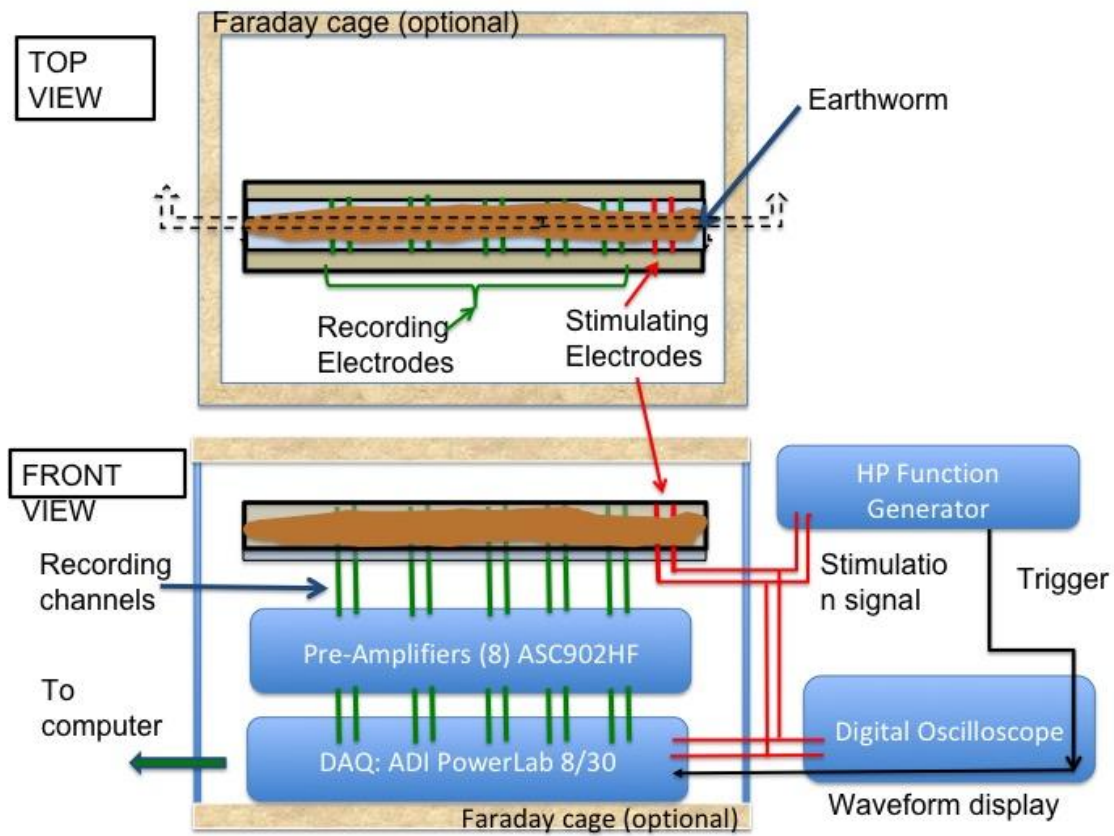

Figure 1. Prototype of testbed (motion-recording camera not shown)

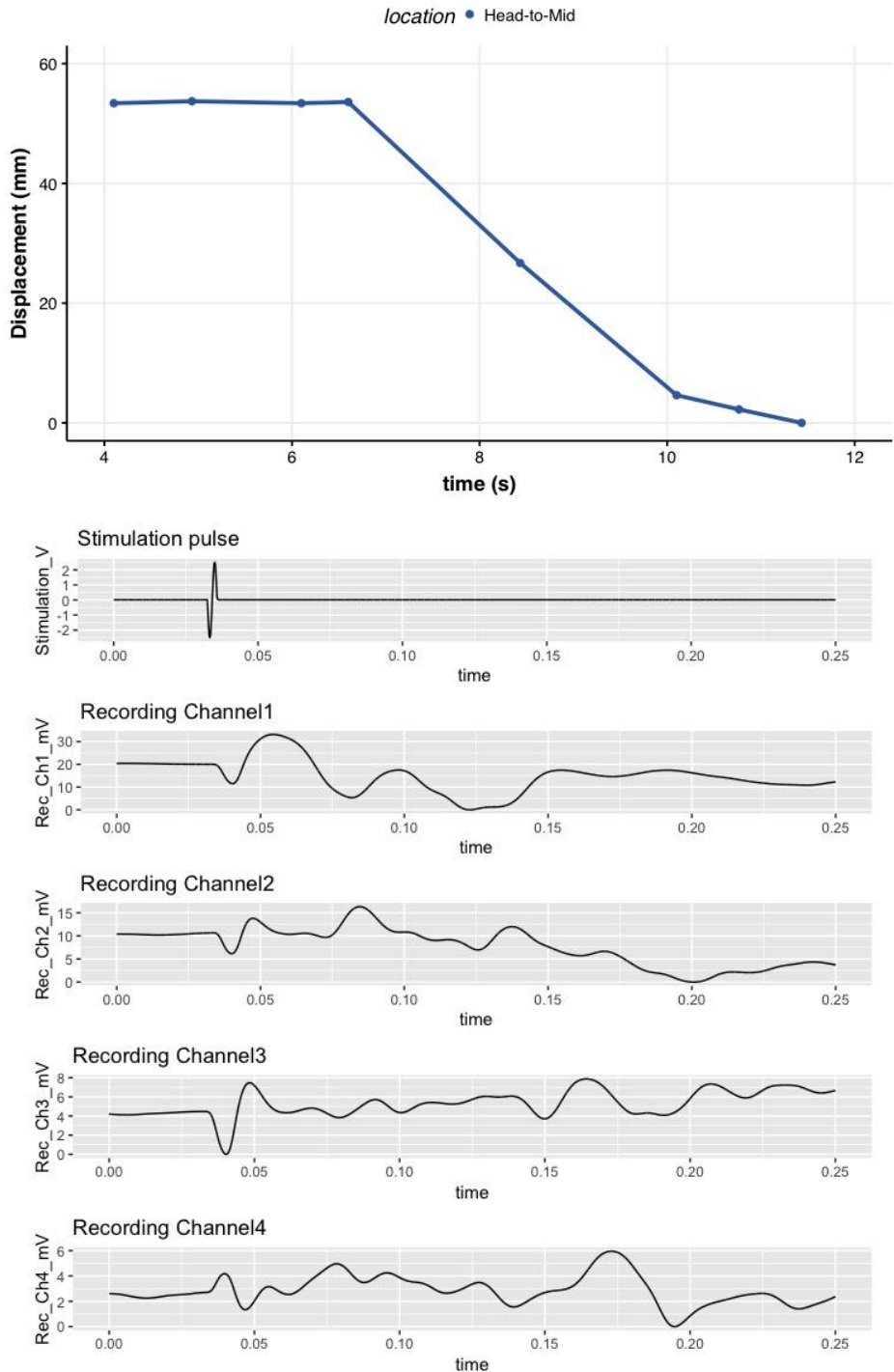

Figure 2. Burst (1 pulse/second) bipolar stimulation with pulse width of 3.3ms, 5Vp-p. Top: Earthworm motion tracking (head-to-midsection) over multiple stimulation pulses. Bottom: Snapshot of action potential recordings for a single stimulation pulse (plots of stimulation pulse and the 4 recording channels are shown).

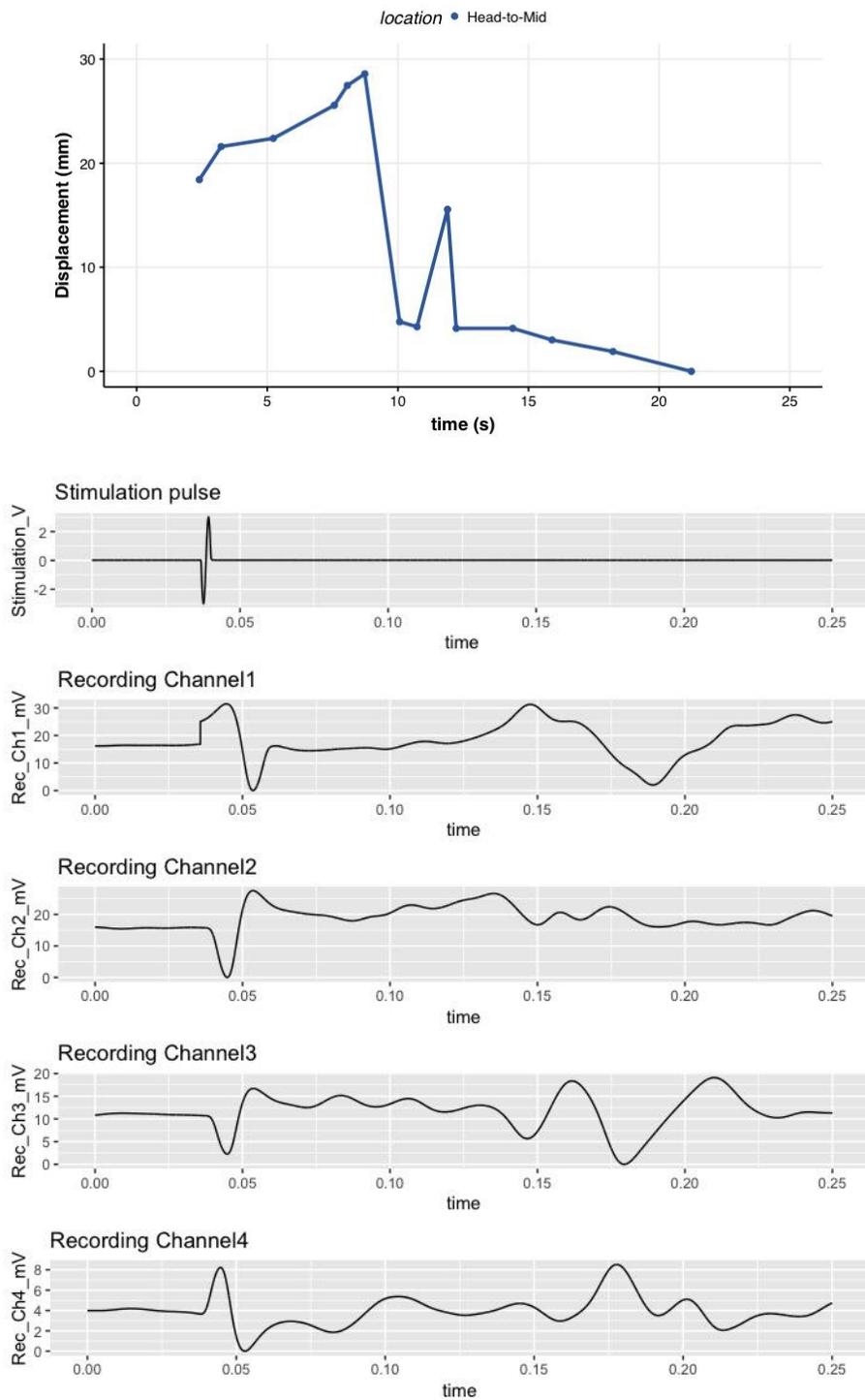

Figure 3. Burst (1 pulse/second) bipolar stimulation with pulse width of 3.3ms, 6Vp-p. Top: Earthworm motion tracking (head-to-midsection) over multiple stimulation pulses. Bottom: Snapshot of action potential recordings for a single stimulation pulse (plots of stimulation pulse and the 4 recording channels are shown).

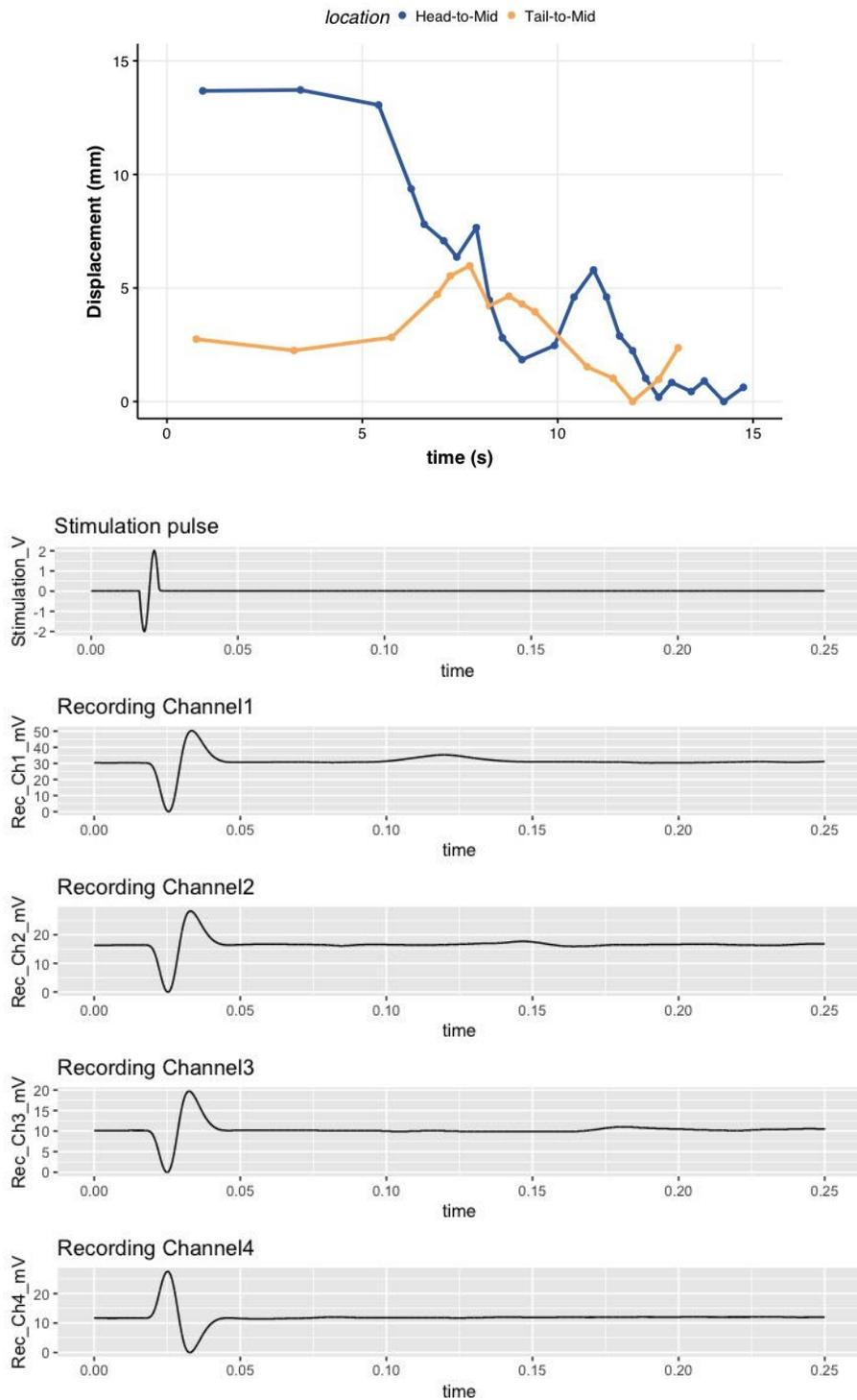

Figure 4. Burst (1 pulse/second) bipolar stimulation with pulse width of 6.67ms, 4Vp-p. Top: Earthworm motion tracking (head-to-midsection and tail-to-midsection) over multiple stimulation pulses. Bottom: Snapshot of action potential recordings for a single stimulation pulse (plots of stimulation pulse and the 4 recording channels are shown).

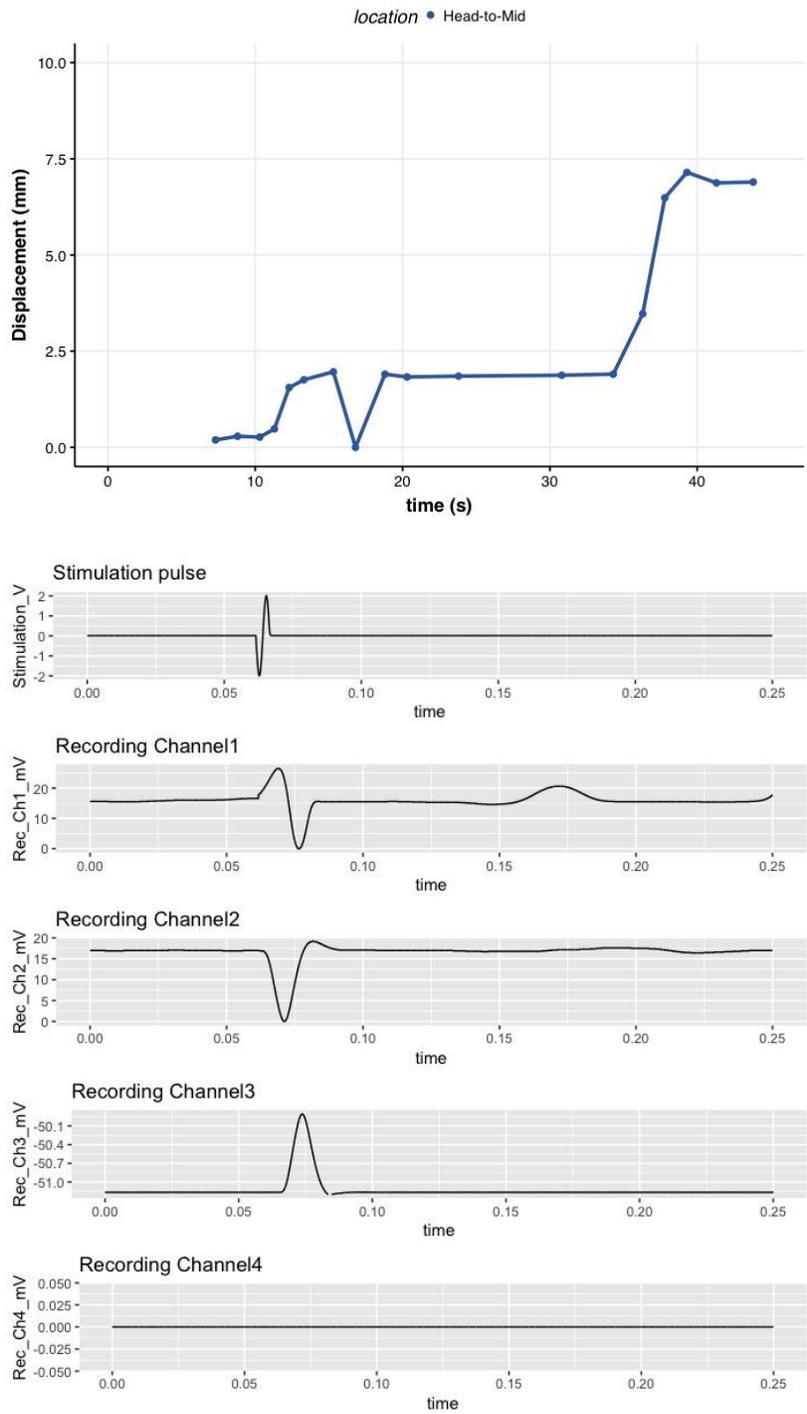

Figure 5. Burst (1 pulse/second) bipolar stimulation with pulse width of 5ms, 4Vp-p. Top: Earthworm motion tracking (head-to-midsection) over multiple stimulation pulses. Bottom: Snapshot of action potential recordings for a single stimulation pulse (plots of stimulation pulse and the 4 recording channels are shown).

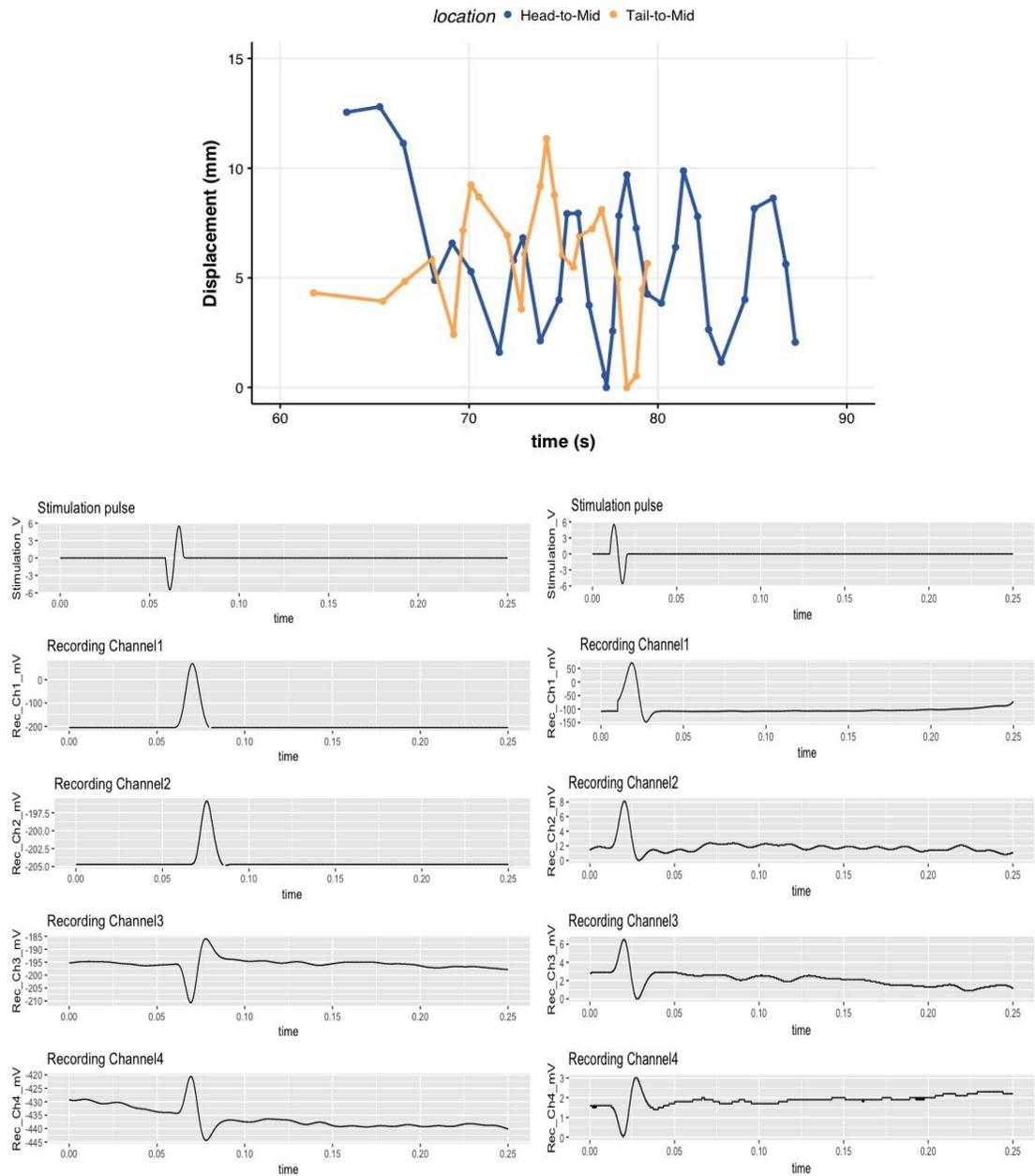

Figure 6. Burst (1 pulse/second) bipolar stimulation with pulse width of 10 ms, 4Vp-p. Top: Earthworm motion tracking (head-to-center and center-to-tail) over multiple stimulation pulses. Bottom left: Snapshot of action potential recordings with normal stimulation (timepoint: 68s). Bottom right: Snapshot of action potential recordings with complement stimulation (timepoint: 98s). All recordings shown for a single stimulation (plots of stimulation pulse and the 4 recording channels are shown).

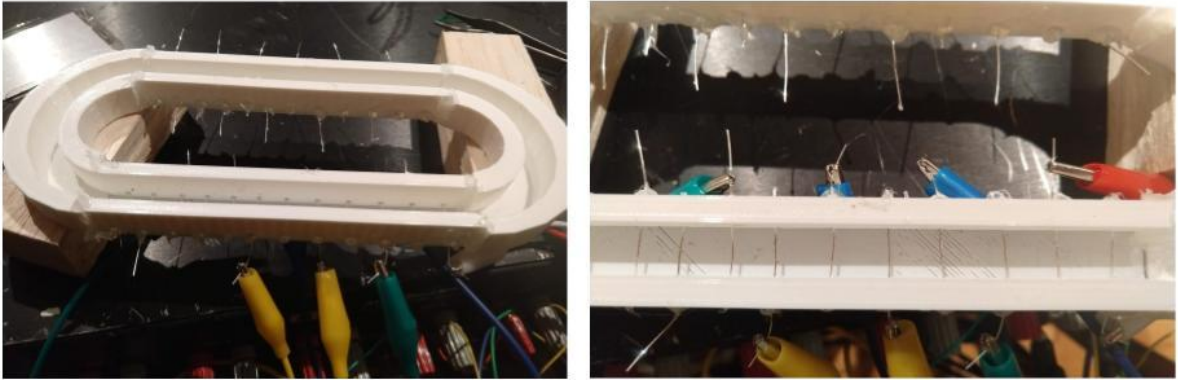

Figure 7. Second-generation 3D-printed experimental rig: (L) side view and (R) top view. Note the silver recording electrodes embedded in the bottom of the raceway.

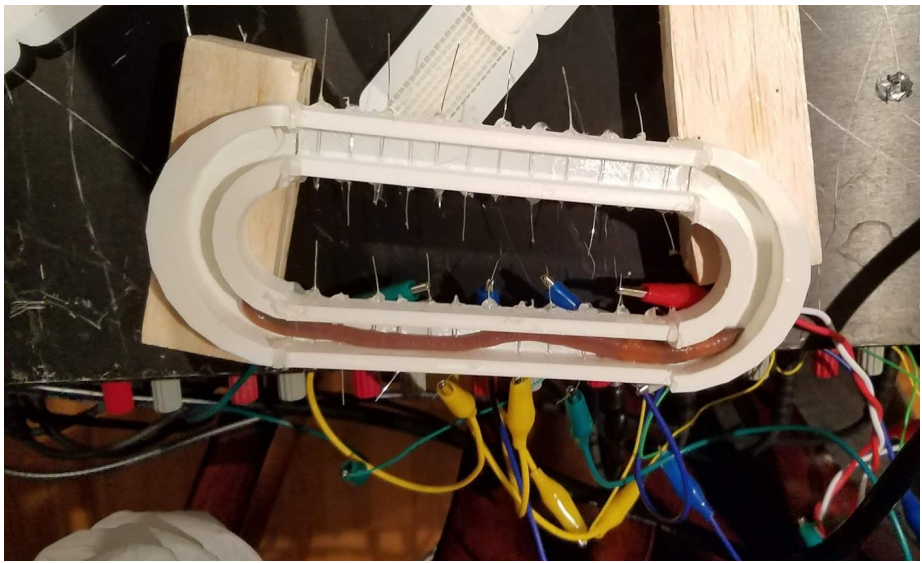

Figure 8. Cold earthworm resting on the silver recording electrodes in the 3D-printed rig.